\documentclass[seceq,preprint]{ptptex}
\usepackage{wrapft}
\usepackage{graphicx}

\newcommand{\bra}[1]{\langle {#1} |}     
\newcommand{\ket}[1]{| {#1} \rangle}     
\newcommand{\wtilde}[1]{\widetilde{#1}} 

\def\l{\left}
\def\r{\right}

\def\<{\langle}
\def\>{\rangle}

\title{
Note on the Time Dependent Variational Approach with Quasi-Spin Squeezed 
State for Pairing Model
}

\author{
Yasuhiko {\sc Tsue}$^1$ and Hideaki {\sc Akaike}$^2$
}

\inst{
$^1$Physics Division, Faculty of Science, Kochi University, Kochi 780-8520, 
Japan\\
$^2$Department of Applied Science, Kochi University, Kochi 780-8520, 
Japan
}

\recdate{
\today
}

\abst{
A simple many-fermion system in which there exists 
$N$ identical fermions in a single spherical orbit with pairing interaction 
is treated by means of the time-dependent variational approach with 
a quasi-spin squeezed state with the aim of going beyond the time-dependent 
Hartree-Fock or Bogoliubov theory. It is shown that the ground state energy 
is reproduced well analytically in this approach. 
}
\begin{document}
\maketitle

\section{Introduction and Preliminary}

The time-dependent variational approach with the $su(2)$-coherent state 
leads to the time-dependent Hartree-Fock (TDHF) or Bogoliubov (TDHB) 
theory for simple many-fermion systems governed by the dynamical 
$su(2)$-algebra.
In the previous paper,\cite{TA04} with the aim of going beyond the 
TDHF approximation, 
we have investigated the dynamics of the Lipkin model by means of 
the time-dependent variational approach with a quasi-spin squeezed state. 
The quasi-spin squeezed state is a possible extension of the $su(2)$-coherent 
state.\cite{TKY94,TAKY96}
Thus, the use of this extended trial state leads to an extended TDHF 
theory.\cite{KPTY01}
In Ref.\citen{TA04}, 
the role of the quantum effects included in the quasi-spin 
squeezed state was analyzed in the Lipkin model. 
Also, in Ref.\citen{ATN04}, the fermionic squeezed state was introduced for 
the $O(4)$ model with a schematic pairing plus quadrapole interaction and 
the role of quantum effects were analyzed. 
The use of the squeezed state in the time-dependent variational approach 
is one of possible candidates to go beyond the TDHF or TDHB approximation. 
However, in the previous works, we have only investigated the role of 
the fluctuations included in the squeezed state based on numerical 
results such as 
the ground state energy and dynamical motion. 
Thus, it is desirable to give an investigation based on an analytical results, 
especially, 
for the approximate reproduction of the ground state energy in dynamical 
viewpoint in order to clarify the role of quantum effects included 
in the quasi-spin squeezed state all the more. 

In this paper, paying an attention to the role of the quantum fluctuation 
included in the quasi-spin squeezed state and interesting to the dynamics 
of a system, 
we treat a simple many-fermion system in which there exists 
$N$ identical fermions in a single spherical orbit with pairing interaction. 
The exact energy eigenvalue of the pairing model is known analytically. 
Thus, we can compare the result obtained in our squeezed state approach 
with the exact one. 
The main aim of this paper is to give an analytical understanding for 
the role of the quantum effects included in the quasi-spin squeezed state. 
Especially, 
the dynamics of the variable representing quantum fluctuations is 
taken into account. 
As a result, it is shown that, under a certain condition, 
the ground state energy is reproduced well analytically, 
like $1/N$ expansion method. 

The single particle state is specified by a set of quantum number 
$(j,m)$, where $j$ and $m$ represent the magnitude of angular momentum of 
the single particle state and 
its projection to the $z$-axis, respectively. 
Let us start with the following Hamiltonian: 
\begin{equation}\label{1-1}
{\hat H}=\epsilon\sum_m{\hat c}^\dagger_{m}{\hat c}_m-\frac{G}{4}
\sum_{m}(-)^{j-m}{\hat c}_m^{\dagger}{\hat c}_{-m}^{\dagger}
\sum_{m'}(-)^{j-m'}{\hat c}_{-m'}{\hat c}_{m'}\ ,
\end{equation}
where $\epsilon$ and $G$ represent the single particle energy and 
the force strength, respectively. The operators ${\hat c}_m$ and 
${\hat c}_m^{\dagger}$ are the fermion annihilation and 
creation operators with the quantum number 
$m$, which obey the anti-commutation relations: 
$
\{\ {\hat c}_m\ , \ {\hat c}_{m'}^{\dagger}\ \}=\delta_{mm'}
$
and 
$\{\ {\hat c}_m\ , \ {\hat c}_{m'}\ \}=
\{\ {\hat c}_m^{\dagger}\ , \ {\hat c}_{m'}^{\dagger}\ \}=0
$.
We introduce the following operators: 
\begin{equation}\label{1-3}
{\hat S}_+=\frac{1}{2}
\sum_{m}{\hat c}_m^{\dagger}{\hat c}_{{\wtilde m}}^{\dagger} \ , \quad
{\hat S}_-=\frac{1}{2}
\sum_{m}{\hat c}_{{\wtilde m}}{\hat c}_{m} \ , \quad
{\hat S}_0=\frac{1}{2}(\sum_m{\hat c}^\dagger_{m}{\hat c}_m-\Omega)\ , 
\end{equation}
where ${\hat c}_{{\wtilde m}}=(-)^{j-m}{\hat c}_{-m}$ and 
$\Omega$ represents the half of the degeneracy: $\Omega=j+1/2$. 
These operators compose the $su(2)$-algebra: 
\begin{equation}\label{1-4}
[{\hat S}_+ , {\hat S}_-]=2{\hat S}_0 \ , \qquad
[{\hat S}_0 , {\hat S}_{\pm}]=\pm{\hat S}_{\pm} \ .
\end{equation}
Thus, these operators are called the quasi-spin operators.\cite{K61,LM65} 
Then, the Hamiltonian (\ref{1-1}) can be rewritten in terms of 
the quasi-spin operators as 
\begin{eqnarray}\label{1-5}
{\hat H}&=&2\epsilon({\hat S}_0+S_j)-G{\hat S}_+{\hat S}_- 
=\epsilon {\hat N}-G{\hat S}_+{\hat S}_- \ , \\
& &\ S_j=\Omega/2\ (=(j+1/2)/2)\ , \nonumber
\end{eqnarray}
where ${\hat N}$ represents the number operator: 
\begin{equation}\label{1-6}
{\hat N}=\sum_{m}{\hat c}_m^{\dagger}{\hat c}_m\ .
\end{equation}
As is well known, the eigenstates and eigenvalues for this Hamiltonian 
are easily obtained. 
Then, the ground state energy can be obtained by setting the seniority 
number being zero:\cite{EG} 
\begin{equation}\label{1-7}
E_0=\epsilon N-\frac{1}{4}GN\Omega
\left(2-\frac{N}{\Omega}+\frac{2}{\Omega}\right)\ . 
\end{equation}

Next, we review the coherent state approach to this pairing model, 
which is identical with the BCS approximation to the pairing model 
consisting of the single energy level. 
The $su(2)$-coherent state is given as 
\begin{eqnarray}\label{1-8}
& &\ket{\phi(\alpha)}=\exp (f{\hat S}_+-f^*{\hat S}_-)\ket{0} \ , \nonumber\\
& &\ket{0}=\ket{S=S_j, S_0=-S_j}\ , \qquad ({\hat S}_-\ket{0}=0)\ .
\end{eqnarray}
We impose the canonicity condition\cite{YK87}:
\begin{equation}\label{1-9}
\bra{\phi(\alpha)} \frac{\partial}{\partial \xi} \ket{\phi(\alpha)}
=\frac{1}{2}\xi^* \ , \qquad
\bra{\phi(\alpha)} \frac{\partial}{\partial \xi^*} \ket{\phi(\alpha)}
=-\frac{1}{2}\xi \ . 
\end{equation}
A possible solution of the above canonicity condition is presented as 
\begin{equation}\label{1-10}
\xi=\sqrt{2S_j}\frac{f}{|f|}\sin |f| \ , \qquad
\xi^*=\sqrt{2S_j}\frac{f^*}{|f|}\sin |f| \ .
\end{equation}
Thus, the expectation values of Hamiltonian ${\hat H}$ 
and number operator ${\hat N}$ are easily calculated and given as 
\begin{eqnarray}\label{1-11}
& &\bra{\phi(\alpha)}{\hat H}\ket{\phi(\alpha)}
=\epsilon \cdot 2|\xi|^2-G\left(2S_j|\xi|^2-
\left(1-\frac{1}{2S_j}\right)|\xi|^4\right)
\equiv E \ , \nonumber\\
& &\bra{\phi(\alpha)}{\hat N}\ket{\phi(\alpha)}=2|\xi|^2 \equiv N \ .
\end{eqnarray}
If total particle number conserves, that is, $N=$constant, then, 
the energy expectation value $E$ is obtained as a function of $N$ : 
\begin{equation}\label{1-12}
E=\epsilon N-\frac{1}{4}GN\Omega\left(2-\frac{N}{\Omega}
+\frac{N}{\Omega^2}\right) \ .
\end{equation}
Compared the approximate ground state energy in (\ref{1-12}) with 
the exact energy eigenvalue in (\ref{1-7}), the last terms in the 
parenthesis of the right-hand side are not identical, which has order of 
$1/\Omega$ if the order of the magnitudes of $N$ and $\Omega$ is 
comparable. 
The main aim of this paper is to show that the last term, 
$2/\Omega$ in (\ref{1-7}), can be recovered by taking into account of 
the dynamics in the time-dependent variational approach with the 
quasi-spin squeezed state.

\section{Quasi-spin squeezed state for pairing model}

In this section, the quasi-spin squeezed state is introduced following to 
Ref.\citen{TKY94}. 
First, we introduce the following operators:
\begin{equation}\label{2-1}
{\hat A}^{\dagger}=\frac{{\hat S}_+}{\sqrt{2S_j}} \ , \qquad
{\hat A}=\frac{{\hat S}_-}{\sqrt{2S_j}} \ , \qquad
{\hat N}=2({\hat S}_0+S_j) \ , 
\end{equation}
where ${\hat N}$ is identical with the number operator (\ref{1-6}). 
Then, the commutation relations can be expressed as 
\begin{equation}\label{2-2}
[{\hat A} , {\hat A}^{\dagger}]=1-\frac{{\hat N}}{2S_j} \ , \qquad
[{\hat N} , {\hat A}]=-2{\hat A} \ , \qquad
[{\hat N} , {\hat A}^{\dagger}]=2{\hat A}^{\dagger} \ . 
\end{equation}
Using the boson-like operator ${\hat A}^{\dagger}$, 
the $su(2)$-coherent state in (\ref{1-8}) can be recast into 
\begin{eqnarray}\label{2-3}
& &\ket{\phi(\alpha)}
=\frac{1}{\sqrt{\Phi(\alpha^*\alpha)}}
\exp(\alpha{\hat A}^{\dagger})\ket{0} \ , \qquad
\Phi(\alpha^*\alpha)=\left(1+\frac{\alpha^*\alpha}{2S_j}\right)^{2S_j} \ , 
\end{eqnarray}
where $\alpha$ is related to $f$ in (\ref{1-8}) as 
$\alpha=\sqrt{2S_j}\cdot (f/|f|)\cdot \tan |f|$. 
The state $\ket{\phi(\alpha)}$ in (\ref{2-3}) is a vacuum state 
for the Bogoliubov-transformed operator ${\hat a}_m$ :
\begin{equation}\label{2-4}
\hat{a}_m=U\hat{c}_m-V(-)^{j-m}\hat{c}^{\dagger}_{-m} \ , 
\qquad \hat{a}_m\ket{\phi(\alpha)}=0 \ .
\end{equation}
The coefficients $U$ and $V$ are given as 
\begin{equation}\label{2-5}
U=\frac{1}{\sqrt{1+\alpha^*\alpha/(2S_j)}} \ , \quad
V=\frac{\alpha}{\sqrt{2S_j}}\frac{1}{\sqrt{1+\alpha^*\alpha/(2S_j)}} \ , \quad
U^2+|V|^2=1 \ . 
\end{equation}
Of course, ${\hat a}_m$ and ${\hat a}_m^{\dagger}$ are fermion annihilation 
and creation operators and the anti-commutation relations are satisfied. 
By using the above Bogoliubov-transformed operators, we introduce 
the following operators: 
\begin{equation}\label{2-6}
{\hat B}^{\dagger}
=\frac{1}{\sqrt{8S_j}}
\sum_m {\hat a}_m^{\dagger}{\hat a}_{{\wtilde m}}^{\dagger} \ , \quad
{\hat B}
=\frac{1}{\sqrt{8S_j}}
\sum_m {\hat a}_{{\wtilde m}}{\hat a}_{m} \ , \quad
{\hat M}=\sum_m{\hat a}_m^{\dagger}{\hat a}_m \ .
\end{equation}
Then, the state $\ket{\phi(\alpha)}$ satisfies 
\begin{equation}\label{2-7}
{\hat B}\ket{\phi(\alpha)}={\hat M}\ket{\phi(\alpha)}=0 \ .
\end{equation}
Further, the commutation relations are as follows:
\begin{equation}\label{2-8}
[{\hat B} , {\hat B}^{\dagger}]=1-\frac{{\hat M}}{2S_j} \ , \qquad
[{\hat M} , {\hat B}]=-2{\hat B} \ , \qquad
[{\hat M} , {\hat B}^{\dagger}]=2{\hat B}^{\dagger} \ . 
\end{equation}
The quasi-spin squeezed state can be constructed on the $su(2)$-coherent 
state $\ket{\phi(\alpha)}$ by using the above 
boson-like operator ${\hat B}^{\dagger}$ as is similar to the ordinary 
boson squeezed state: 
\begin{eqnarray}\label{2-9}
& &\ket{\psi(\alpha,\beta)}
=\frac{1}{\sqrt{\Psi(\beta^*\beta)}}\exp \left(
\frac{1}{2}\beta{\hat B}^{\dagger 2}\right) \ket{\phi(\alpha)} \ , \nonumber\\
& &\Psi(\beta^*\beta)=1+\sum_{k=1}^{[S_j]}\frac{(2k-1)!!}{(2k)!!}
\prod_{p=1}^{2k-1}\left(1-\frac{p}{2S_j}\right)(|\beta|^2)^k \ .
\end{eqnarray}
We call the state $\ket{\psi(\alpha,\beta)}$ the quasi-spin squeezed state.

We can easily 
calculate the expectation values for various operators with respect 
to the quasi-spin squeezed state. 
The expectation values can be expressed in terms of the canonical 
variables which are introduced through the canonicity conditions. 
The same results derived in the following are originally given 
in the Lipkin model in Ref.\citen{Y94} at the first time 
and these results were used in Ref.\citen{TA04} to analyze the effects 
of quantum fluctuations in the Lipkin model. 
For the quasi-spin squeezed state $\ket{\psi(\alpha,\beta)}$ in (\ref{2-9}), 
the following expression is useful : 
\begin{eqnarray}\label{2-10}
\bra{\psi(\alpha,\beta)}\partial_z \ket{\psi(\alpha,\beta)}
&=&\frac{\Psi'(\beta^*\beta)}{\Psi(\beta^*\beta)}\frac{1}{2}
(\beta^*\partial_z \beta - \beta \partial_z \beta^*) \nonumber\\
& &+\left(1-\frac{2\beta^*\beta}{S_j}
\frac{\Psi'(\beta^*\beta)}{\Psi(\beta^*\beta)}\right)
\cdot \frac{1}{1+\alpha^*\alpha/(2S_j)}\frac{1}{2}
(\alpha^* \partial_z \alpha - \alpha \partial_z \alpha^*)\ , \nonumber\\
\Psi'(u)&\equiv&\frac{\partial \Psi(u)}{\partial u} \ , 
\end{eqnarray}
where $\partial_z=\partial/\partial z$. 
The canonicity conditions are imposed in order to introduce the sets of 
canonical variables $(X, X^*)$ and $(Y, Y^*)$ as follows :
\begin{eqnarray}\label{2-11}
& &\bra{\psi(\alpha,\beta)}\partial_X \ket{\psi(\alpha,\beta)}
=\frac{1}{2}X^* \ , \qquad
\bra{\psi(\alpha,\beta)}\partial_{X^*} \ket{\psi(\alpha,\beta)}
=-\frac{1}{2}X \ , \nonumber\\
& &\bra{\psi(\alpha,\beta)}\partial_Y \ket{\psi(\alpha,\beta)}
=\frac{1}{2}Y^* \ , \qquad
\bra{\psi(\alpha,\beta)}\partial_{Y^*} \ket{\psi(\alpha,\beta)}
=-\frac{1}{2}Y \ . 
\end{eqnarray}
Possible solutions for $X$ and $Y$ are obtained as 
\begin{eqnarray}\label{2-12}
& &\alpha=\frac{X}{\sqrt{1-\frac{X^*X}{2S_j}-\frac{4Y^*Y}{2S_j}}} \ , \quad
\alpha^*=\frac{X^*}{\sqrt{1-\frac{X^*X}{2S_j}-\frac{4Y^*Y}{2S_j}}} \ ,
\nonumber\\
& &\beta=\frac{Y}{\sqrt{K(Y^*Y)}} \ , \qquad
\beta^*=\frac{Y^*}{\sqrt{K(Y^*Y)}} \ , 
\end{eqnarray}
where $K(Y^*Y)$ is introduced and satisfies the relation 
\begin{equation}\label{2-13}
K(Y^*Y)\Psi(Y^*Y/K)=\Psi'(Y^*Y/K) \ .
\end{equation}
The expectation values for ${\hat B}$, ${\hat B}^{\dagger}$, 
${\hat M}$ and the products of these operators are easily obtained 
and are expressed in terms of the canonical variables as 
follows : 
\begin{subequations}\label{2-14}
\begin{eqnarray}
& &\bra{\psi(\alpha,\beta)}{\hat B}\ket{\psi(\alpha,\beta)}
=\bra{\psi(\alpha,\beta)}{\hat B}^{\dagger}\ket{\psi(\alpha,\beta)}=0 \ ,
\nonumber\\
& &\bra{\psi(\alpha,\beta)}{\hat M}\ket{\psi(\alpha,\beta)}
=4Y^*Y \ , 
\label{2-14a}\\
& &\bra{\psi(\alpha,\beta)}{\hat B}^2\ket{\psi(\alpha,\beta)}
=2Y\sqrt{K(Y^*Y)} \ , 
\nonumber\\
& &
\bra{\psi(\alpha,\beta)}{\hat B}^{\dagger 2}\ket{\psi(\alpha,\beta)}
=2Y^*\sqrt{K(Y^*Y)} \ , 
\label{2-14b}\\
& &\bra{\psi(\alpha,\beta)}{\hat B}^{\dagger}{\hat B}\ket{\psi(\alpha,\beta)}
=2(1-1/(2S_j))Y^*Y-\frac{2}{S_j}(Y^*Y)^2\cdot L(Y^*Y) \ , \nonumber\\
& &\bra{\psi(\alpha,\beta)}{\hat B}{\hat B}^{\dagger}\ket{\psi(\alpha,\beta)}
=\bra{\psi(\alpha,\beta)}{\hat B}^{\dagger}{\hat B}\ket{\psi(\alpha,\beta)}
+(1-2Y^*Y/S_j) \ , 
\label{2-14c}\\
& &\bra{\psi(\alpha,\beta)}{\hat M}^2\ket{\psi(\alpha,\beta)}
=16Y^*Y(1+Y^*Y\cdot L(Y^*Y)) \ , 
\label{2-14d}
\end{eqnarray}
where $L(Y^*Y)$ is defined and satisfies 
\begin{equation}\label{2-14e}
K(Y^*Y)^2\cdot L(Y^*Y)=\frac{\Psi''(\beta^*\beta)}{\Psi(\beta^*\beta)} \ .
\end{equation}
\end{subequations}

By using the relations between the original variables $\alpha$ and $\beta$ 
and the canonical variables $X$ and $Y$, 
the coefficients of the Bogoliubov transformation (\ref{2-5}), 
$U$ and $V$, are expressed as 
\begin{equation}\label{2-15}
U=\frac{\sqrt{1-\frac{X^*X}{2S_j}-\frac{4Y^*Y}{2S_j}}}
{\sqrt{1-\frac{4Y^*Y}{2S_j}}} \ , \quad
V=\frac{X}{\sqrt{2S_j}}\frac{1}{\sqrt{1-\frac{4Y^*Y}{2S_j}}} \ .
\end{equation}
Then, the operators ${\hat A}$, ${\hat A}^{\dagger}$ and ${\hat N}$, 
which are related to the quasi-spin operators ${\hat S}_-$, ${\hat S}_+$ 
and ${\hat S}_0$, respectively, in (\ref{2-1}), can be 
expressed as 
\begin{eqnarray}\label{2-16}
& &{\hat A}=\sqrt{2S_j}UV\left(1-\frac{\hat M}{2S_j}\right)
-V^2{\hat B}^{\dagger}+U^2{\hat B} \ , \nonumber\\
& &{\hat A}^{\dagger}=\sqrt{2S_j}UV^*\left(1-\frac{\hat M}{2S_j}\right)
+U^2{\hat B}^{\dagger}-V^{*2}{\hat B} \ , \nonumber\\
& &{\hat N}=4S_jV^*V\left(1-\frac{\hat M}{2S_j}\right)+\sqrt{2S_j}U
(V{\hat B}^{\dagger} +V^*{\hat B})+{\hat M} \ . 
\end{eqnarray}
Thus, the expectation values for ${\hat A}$, ${\hat A}^{\dagger}$, 
${\hat N}$ and the products of these operators are easily obtained 
and are expressed in terms of the canonical variables. 
For example, from (\ref{2-1}), the expectation values of quasi-spin operators 
are derived as 
\begin{eqnarray}\label{2-18}
& &\bra{\psi(\alpha,\beta)}{\hat S}_+\ket{\psi(\alpha,\beta)}
=X^*\sqrt{2S_j-X^*X-4Y^*Y} \ , \nonumber\\
& &\bra{\psi(\alpha,\beta)}{\hat S}_-\ket{\psi(\alpha,\beta)}
=\sqrt{2S_j-X^*X-4Y^*Y}\ X \ , \nonumber\\
& &\bra{\psi(\alpha,\beta)}{\hat S}_0\ket{\psi(\alpha,\beta)}
=X^*X+2Y^*Y-S_j
\end{eqnarray}
and also the expectation value of the number operator in (\ref{1-6}) is 
calculated as 
\begin{equation}\label{2-18add}
N=\bra{\psi(\alpha,\beta)}{\hat N}\ket{\psi(\alpha,\beta)}
=2X^*X+4Y^*Y \ . 
\end{equation}
The above expressions in (\ref{2-18}) 
correspond to the Holstein-Primakoff boson 
realization for the $su(2)$-algebra such as 
$S_+=X^*\sqrt{2S_j-X^*X}$ and so on. Thus, we can conclude that 
the variable $|Y|^2$ represents the quantum effect.

The model Hamiltonian (\ref{1-5}) can be expressed in terms of 
the fermion number operator ${\hat N}$ and the boson-like operators 
${\hat A}$ and ${\hat A}^{\dagger}$ as 
\begin{equation}\label{2-19}
{\hat H}=\epsilon {\hat N}-2S_j G{\hat A}^{\dagger}{\hat A} \ .
\end{equation}
Thus, the expectation value of this Hamiltonian is easily 
obtained. 
We denote it as $H_{\rm sq}$ : 
\begin{eqnarray}\label{2-20}
H_{\rm sq}&=&\bra{\psi(\alpha,\beta)}{\hat H}\ket{\psi(\alpha,\beta)} 
\nonumber\\ 
&=&\epsilon(2n_{X}+4n_{Y})-2GS_{j}
\l\{
{\cal X}+\frac{n_{X}^{2}}{2S_{j}(2S_{j}-4n_{Y})}
-\frac{4}{S_{j}^{2}}{\cal X}{\cal Y}\l[n_{Y}+n_{Y}^{2} L-n_{Y}^{2} \r]
\r.\nonumber\\
& &\l.
-\frac{2}{S_{j}}\sqrt{n_{Y}K}{\cal X}{\cal Y}
\cos(2\theta_{X}-\theta_{Y})
+2\l[
1-\frac{{\cal X}{\cal Y}}{S_{j}}
\r]
\l[
\l(
1-\frac{1}{2S_{j}}
\r)n_{Y}
-\frac{n_{Y}^{2}\cdot L}{S_{j}}
\r]
\r\}\ , \nonumber\\
& &
\end{eqnarray}
where we introduce the action-angle variables instead of 
$(X,X^*)$ and $(Y, Y^*)$ as 
\begin{eqnarray}\label{2-21}
& &X=\sqrt{n_X}e^{-i\theta_X} \ , \qquad X^*=\sqrt{n_X}e^{i\theta_X} \ , 
\nonumber\\ 
& &Y=\sqrt{n_Y}e^{-i\theta_Y} \ , \qquad Y^*=\sqrt{n_Y}e^{i\theta_Y}  
\end{eqnarray}
and ${\cal X}$ and ${\cal Y}$ are defined as 
\begin{eqnarray}\label{2-22}
{\cal X}=\l(
n_{X}-\frac{n_{X}^{2}}{2S_{j}}-\frac{2n_{Y}n_{X}}{S_{j}}
\r),~~~~~
{\cal Y}=\frac{1}{\l(1-\displaystyle\frac{2n_{Y}}{S_{j}}\r)^{2}}.
\end{eqnarray}

The dynamics of this system can be investigated approximately by determining 
the time-dependence of the canonical variables $(X, X^*)$ and $(Y, Y^*)$ 
or $(n_X, \theta_X)$ and $(n_Y, \theta_Y)$. 
The time-dependence of these canonical variables is derived from the 
time-dependent variational principle : 
\begin{equation}\label{2-23}
\delta\int \bra{\psi(\alpha,\beta)}i\partial_t-{\hat H}
\ket{\psi(\alpha,\beta)}dt=0 \ .
\end{equation}

\section{Time evolution of variational state}

In the $su(2)$-coherent state approximation, 
the expectation value of the Hamiltonian 
is calculated as 
\begin{eqnarray}\label{3-1}
H_{\rm ch}=\bra{\phi(\alpha)}\hat{H}\ket{\phi(\alpha)}
=
2\epsilon~\!n_{X}-2GS_{j}
\l\{
n_{X}-\frac{n_{X}}{2S_{j}}+\frac{n_{X}^{2}}{4S_{j}^{2}}
\r\} \ .
\end{eqnarray}
In this approximation, namely, usual time-dependent 
Hartree-Bogoliubov 
approximation, the canonical equations of motion derived from 
the time-dependent variational principle have the following forms: 
\begin{eqnarray}\label{3-2}
& &\dot{\theta_{X}}=\frac{\partial H_{\rm ch}}{\partial n_{X}}=2\epsilon
-2GS_{j}
\l(1-\frac{n_{X}}{S_{j}}-\frac{n_{X}}{2S_{j}^{2}} \r)\ , \nonumber\\
& &\dot{n_{X}}=-\frac{\partial H_{\rm ch}}{\partial \theta_{X}}
=0 \ .
\end{eqnarray}
The solutions of the above equations of motion are easily obtained as 
\begin{eqnarray}\label{3-3}
& &\theta_{X}(t)=
2\l[
\epsilon-GS_{j}\l(1-\frac{n_{0}}{S_{j}}-\frac{n_{0}}{2S_{j}^{2}} \r)
\r]~\!t+\theta_{X0}\ , \nonumber\\
& &n_{X}(t)=n_{0}~({\rm constant}) \ .
\end{eqnarray}

\begin{figure}[b]
 \parbox{\halftext}
 {  
  \centerline{\includegraphics[width=7cm]{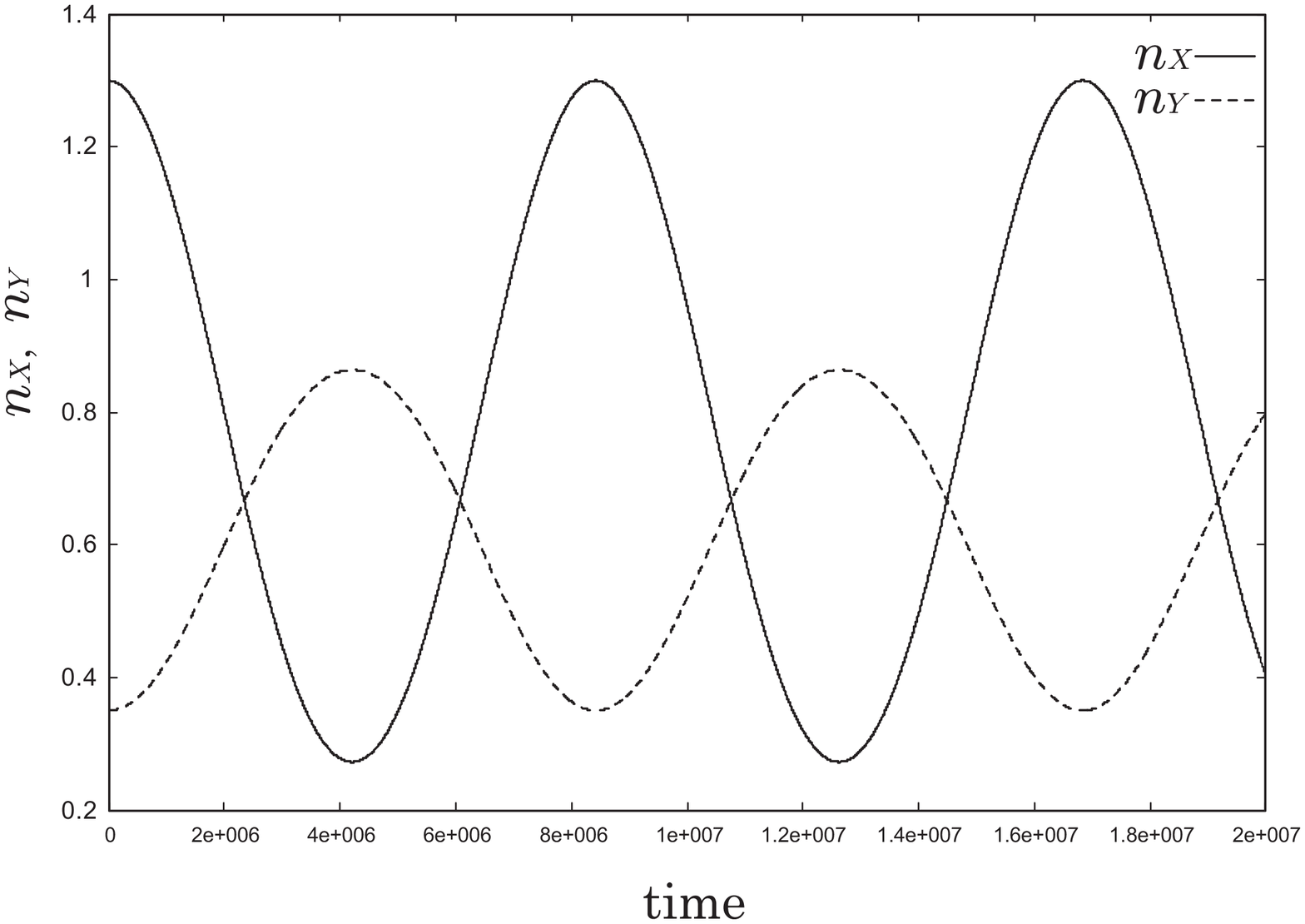}}
  \caption{
  The time evolution of $n_{X}(t)$ and $n_{Y}(t)$ is plotted
  in the case of the   quasi-spin squeezed state approach. 
  The parameters are taken as $G=1.2$, $\epsilon=1.0$ and $\Omega=2N=8$. 
  The initial values are $n_{X}(t=0)=1.3$, $\theta_{X}(t=0)=0.0$, 
  $n_{Y}(t=0)=0.35$ and $\theta_{Y}(t=0)=0.0$. 
  		   }
   \label{fig:Fig1}
 }
 \hspace{3mm}
 \parbox{\halftext}
 {  
  \centerline{\includegraphics[width=7cm]{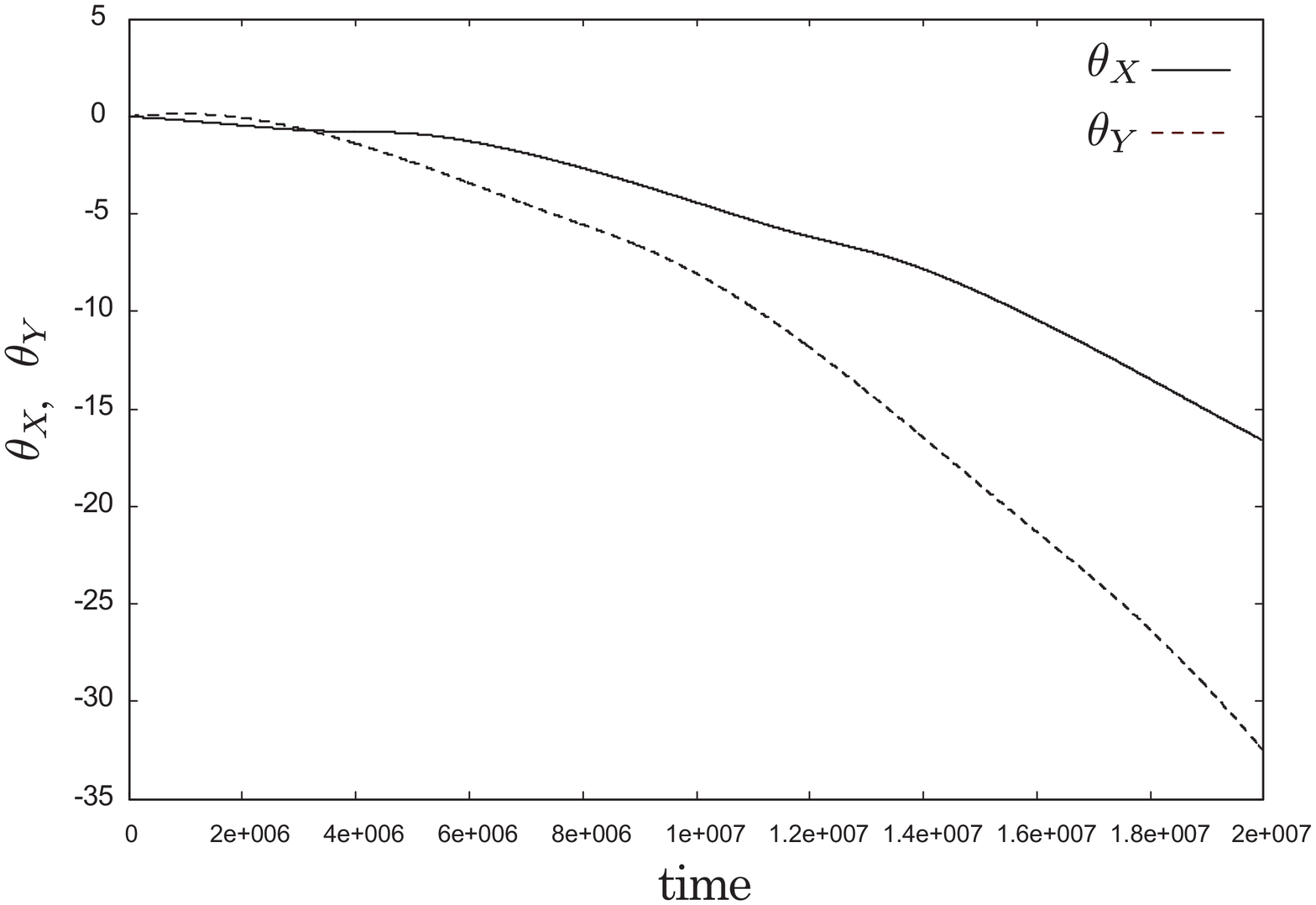}}
  \caption{
  The time evolution of $\theta_{X}(t)$ and $\theta_{Y}(t)$ is plotted
  in the case of the   quasi-spin squeezed state approach. 
  The parameters are taken as $G=1.2$, $\epsilon=1.0$ and $\Omega=2N=8$. 
  The initial values are $n_{X}(t=0)=1.3$, $\theta_{X}(t=0)=0.0$, 
  $n_{Y}(t=0)=0.35$ and $\theta_{Y}(t=0)=0.0$. 
  		   }
 \label{fig:Fig2}
 }
\end{figure}

On the other hand, in the quasi-spin squeezed state approximation, 
the equations of motion derived from (\ref{2-23}) are written as 
\begin{eqnarray}
\dot{\theta_{X}}&=&\frac{\partial H_{\rm sq}}{\partial n_{X}}\nonumber\\
&=&
2\epsilon-2GS_{j}
\l\{
\frac{\partial {\cal X}}{\partial n_{X}}+\frac{n_{X}}{S_{j}(2S_{j}-4n_{Y})}
+\frac{4}{S_{j}^{2}}\cdot\frac{\partial {\cal X}}{\partial n_{X}}\cdot
{\cal Y}\cdot[n_{Y}+n_{Y}^{2} L-n_{Y}^{2} ]
\r.\nonumber\\
& &\qquad\qquad\ \ 
\l. -\frac{2}{S_{j}}\sqrt{n_{Y}K}\cdot
\frac{\partial {\cal X}}{\partial n_{X}}
\cos(2\theta_{X}-\theta_{Y})
-2\cdot\frac{\partial {\cal X}}{\partial n_{X}}
\frac{{\cal Y}}{S_{j}}\cdot L \r\} \ , \nonumber\\
\dot{n_{X}}&=&
-\frac{\partial H_{\rm sq}}{\partial \theta_{X}}
=8G\sqrt{n_{Y}K}~\!{\cal X}{\cal Y}\sin(2\theta_{X}-\theta_{Y}) \ , 
\label{3-4}\\
\dot{\theta_{Y}}&=&
\frac{\partial H_{\rm sq}}{\partial n_{Y}}\nonumber\\
&=&
4\epsilon-2GS_{j}
\l\{
\frac{\partial {\cal X}}{\partial n_{Y}}+
\frac{n_{X}^{2}\cdot {\cal Y}}{2S_{j}^{3}}
~\!+~\!
\frac{4}{S_{j}^{2}}\cdot
\l[
\frac{\partial {\cal X}}{\partial n_{Y}}{\cal Y}
+
\frac{\partial {\cal Y}}{\partial n_{Y}}{\cal X}
\r]
\cdot
{\cal Y}\cdot[n_{Y}+n_{Y}^{2} L-n_{Y}^{2} ]
\r.\nonumber\\
& &\qquad\qquad
+\frac{4}{S_{j}^{2}}\cdot {\cal X}{\cal Y}\cdot
[1+2n_{Y}L+n_{Y}^{2}\frac{\partial L}{\partial n_Y}-2n_{Y}]\nonumber\\
& &\qquad\qquad
-\frac{1}{S_{j}\sqrt{n_{Y}K}}
(K+n_{Y}\frac{\partial K}{\partial n_Y}){\cal X}{\cal Y}
\cos(2\theta_{X}-\theta_{Y})\nonumber\\
& &\qquad\qquad
-\frac{2}{S_{j}}\sqrt{n_{Y}K}\l[
\frac{\partial {\cal X}}{\partial n_{Y}}{\cal Y}
+
\frac{\partial {\cal Y}}{\partial n_{Y}}{\cal X}
\r]\cos(2\theta_{X}-\theta_{Y})\nonumber\\
& &\qquad\qquad
-\frac{2}{S_{j}}
\l[
\frac{\partial {\cal X}}{\partial n_{Y}}{\cal Y}
+
\frac{\partial {\cal Y}}{\partial n_{Y}}{\cal X}
\r]
\cdot
\l[
\l(1-\frac{1}{2S_{j}}\r)n_{Y}
-\frac{n_{Y}^{2}}{S_{j}}L
\r]\nonumber\\
& &\qquad\qquad
\l.+2
\l[
1-\frac{{\cal X}{\cal Y}}{S_{j}}
\r]\cdot
\l[
\l(
1-\frac{1}{2S_{j}}
\r)
-\frac{1}{S_{j}}
\l[
2n_{Y}L+n_{Y}^{2}\frac{\partial L}{\partial n_Y}
\r]
\r]
\r\} \ , \nonumber\\
\dot{n_{Y}}&=&-\frac{\partial H_{sq}}{\partial \theta_{Y}}
=-4G\sqrt{n_{Y}K}~\!{\cal X}{\cal Y}~\!\sin (2\theta_{X}-\theta_{Y})
\label{3-5}
\end{eqnarray}
Here, the number conservation is satisfied: 
\begin{equation}\label{3-6}
{\dot N}=2{\dot n_X}+4{\dot n_Y}=0 \ .
\end{equation}
The time evolution of $n_X(t)$ and $n_Y(t)$ in Fig.\ref{fig:Fig1} and 
$\theta_{X}(t)$ and $\theta_{Y}(t)$ in Fig.\ref{fig:Fig2} is 
plotted with appropriate initial conditions. 
The parameters used here are $G=1.2$, $\epsilon=1.0$ and 
$\Omega=2N=8$. In the $su(2)$-coherent state approach, $n_X$ is constant 
of motion. However, quasi-spin squeezed state approach, $n_X$ and $n_Y$ 
oscillate with antiphase, while the total particle number is conserved.

\section{Dynamical approach to the ground state energy}

Hereafter, we assume that $|Y|^2\ (=n_Y) \ll 1$ 
because $|Y|^2$ means the quantum 
fluctuations. 
Then, $K(Y^*Y)$ and $L(Y^*Y)$ defined in (\ref{2-13}) and (\ref{2-14e}), 
respectively, can be evaluated by the expansion with respect to $|Y|^2$. 
As a result, we obtain 
\begin{eqnarray}\label{2-24}
K(Y^*Y)&=&\frac{1}{2}\left(1-\frac{1}{2S_j}\right)
+\left(1-\frac{7}{2S_j}+\frac{9}{(2S_j)^2}\right)Y^*Y+\cdots \ . 
\nonumber\\
L(Y^*Y)&=&\frac{1}{1-\frac{1}{2S_j}}3\left(1-\frac{2}{2S_j}\right)\!\!
\left(1-\frac{3}{2S_j}\right) \nonumber\\
& &-\frac{48}{2S_j}\frac{1}{(1-\frac{1}{2S_j})^2}
\left(1-\frac{2}{2S_j}\right)\!\!\left(1-\frac{3}{2S_j}\right)\!\!
\left(1-\frac{4}{2S_j}\right)Y^*Y+\cdots \ . \qquad
\end{eqnarray}
Then, the expectation values for the Hamiltonian, the time-derivative and 
the number operator can be expressed as 
\begin{subequations}\label{2-25}
\begin{eqnarray}
& &H_{\rm sq}=
\epsilon(2n_X+4n_Y) -G\biggl[2S_j n_X-n_X^2
+\frac{n_X^2}{2S_j} \nonumber\\
& &\qquad\quad
  -2\sqrt{2}\sqrt{1-\frac{1}{2S_j}}\left(1-\frac{n_X}{2S_j}\right)
n_X\sqrt{n_Y}\cos (2\theta_X-\theta_Y) \nonumber\\
& &\qquad\quad
  +2\left(2S_j-1-4n_X+\frac{10}{2S_j}n_X
  +\frac{2}{2S_j}n_X^2-\frac{8}{(2S_j)^2}n_X^2\right)n_Y 
+{\rm O}(n_Y^{3/2}) \biggl] 
\ , \nonumber\\
& &\label{2-25a}\\
& &\bra{\psi(\alpha,\beta)}i\partial_t\ket{\psi(\alpha,\beta)}
=(n_X{\dot \theta}_X+n_Y{\dot \theta}_Y) \ , 
\label{2-25b}\\
& &N=\bra{\psi(\alpha,\beta)}{\hat N}\ket{\psi(\alpha,\beta)}
=2n_X+4n_Y \ . 
\label{2-25c}
\end{eqnarray}
\end{subequations}
From the time-dependent variational principle (\ref{2-23}) or 
(\ref{3-4}) and (\ref{3-5}), the following 
equations of motion are derived under the above-mentioned approximation : 
\begin{subequations}\label{2-26}
\begin{eqnarray}
& &
{\dot \theta}_X=\frac{\partial H_{\rm sq}}{\partial n_X}
\approx 2\epsilon-G\!\cdot\! 2\biggl[
S_j\!-\!n_X\!+\!\frac{n_X}{2S_j}\!-\!\sqrt{2}\sqrt{1-\frac{1}{2S_j}}
\!\!\left(
1-\frac{n_X}{S_j}\right)\!\!\sqrt{n_Y}\cos (2\theta_X-\theta_Y) \nonumber\\
& &\qquad\qquad\qquad\qquad\qquad
+4\left(-1+\frac{5}{4S_j}+\frac{1}{2S_j}n_X-\frac{1}{S_j^2}n_X\right)
n_Y\biggl] \ , 
\label{2-26a}\\
& &{\dot n}_X=-\frac{\partial H_{\rm sq}}{\partial \theta_X}
\approx -G\cdot 4\sqrt{2}\sqrt{1-\frac{1}{2S_j}}\!\left(1-\frac{n_X}{2S_j}
\right)\!n_X\sqrt{n_Y}\sin (2\theta_X-\theta_Y) \ , 
\label{2-26b}\\
& &
{\dot \theta}_Y=\frac{\partial H_{\rm sq}}{\partial n_Y}
\approx 4\epsilon-G\biggl[
-\sqrt{2}\sqrt{1-\frac{1}{2S_j}}\!\!\left(
1-\frac{n_X}{2S_j}\right)\!\!\frac{n_X}{\sqrt{n_Y}}\cos (2\theta_X-\theta_Y) 
\nonumber\\
& &\qquad\qquad\qquad\qquad\qquad
+2\left(-1+2S_j-4n_X+\frac{5}{S_j}+\frac{1}{S_j}n_X^2-\frac{2}{S_j^2}n_X^2
\right)\biggl] \ , 
\label{2-26c}\\
& &{\dot n}_Y=-\frac{\partial H_{\rm sq}}{\partial \theta_Y}
\approx G\cdot 2\sqrt{2}\sqrt{1-\frac{1}{2S_j}}\!\left(1-\frac{n_X}{2S_j}
\right)\!n_X\sqrt{n_Y}\sin (2\theta_X-\theta_Y) \ . 
\label{2-26d}
\end{eqnarray}
\end{subequations}
It is found from (\ref{2-26b}) and (\ref{2-26d}) that the total 
fermion number $N$ in (\ref{2-25c}) is also conserved in this 
approximation, that is, 
\begin{equation}\label{2-27}
{\dot N}=2{\dot n}_X+4{\dot n}_Y=0 \ . 
\end{equation}

It should be noted here that $n_X \leq N/2$ from (\ref{2-25c}) and 
$N \leq \Omega/2=S_j$. Thus, the inequality $1-n_X/2S_j \geq 0$ 
is obtained. From the approximated energy expectation value (\ref{2-25a}) for 
$G>0$, 
the energy minimum is then obtained in the case $\cos(2\theta_X-\theta_Y)=-1$, 
namely, 
\begin{equation}\label{3-28}
\theta_Y=2\theta_X+\pi \ . 
\end{equation}
Since the energy is minimal in the ground state, 
the relation (\ref{3-28}) should be satisfied at any time. 
In order to assure the above-mentioned situation, the following 
consistency condition should be obeyed : 
\begin{equation}\label{3-29}
{\dot \theta}_Y=2{\dot \theta}_X \ . 
\end{equation}
Thus, from the equations of motion (\ref{2-26a}) and (\ref{2-26c}), 
under the approximation of small $n_Y$, the consistency condition 
(\ref{3-29}) gives the following expression of $n_Y$ 
in the lowest order approximation of $n_Y$ : 
\begin{eqnarray}\label{3-30}
\sqrt{n_Y}&\approx& \frac{\sqrt{1-\frac{1}{2S_j}}}
{2\sqrt{2}(1-\frac{2}{S_j})}\cdot
\left[1+\frac{1}{2n_X}\frac{1}{(1-\frac{n_X}{2S_j})(1-\frac{2}{S_j})}
\right]^{-1} \ .
\end{eqnarray}
Thus, by substituting (\ref{3-28}) and 
$\sqrt{n_Y}$ in (\ref{3-30}) under the lowest order 
approximation of $n_Y$ into 
the energy expectation value (\ref{2-25a}), 
and by performing the approximation 
of large $N$ or large $\Omega (=2S_j)$ approximation, we obtain the 
ground state energy as 
\begin{equation}\label{3-31}
H_{\rm sq}=\epsilon N-\frac{1}{4}GN\Omega
\left(2-\frac{N}{\Omega}+\frac{2}{\Omega} + 
{\rm O}(1/(N\Omega), 1/\Omega^2, 1/N^2)\right) \ . 
\end{equation}
This result reproduces the exact energy eigenvalue (\ref{1-7}) 
by neglecting the higher order term of $1/(N\Omega)$, $1/\Omega^2$ and 
$1/N^2$ for large $N$ and $\Omega$ limit. 
Thus, the quasi-spin squeezed state presents a good approximation 
in the time-dependent variational approach to the pairing model. 
In this approach, the existence of the rotational motion in the phase 
space consisting of $(n_X, \theta_X; n_Y, \theta_Y)$ plays the important 
role. The angle variables for rotational motion, $\theta_X$ and $\theta_Y$, 
are consistently changed in (\ref{3-29}). 
This consistency condition is essential to reproduce the exact 
energy for the ground state under the large $N$ and $\Omega$ limit. 
The approximation corresponds to so-called large $N$ approximation. 
In general, it is known that the large $N$ expansion at zero temperature 
corresponds to $\hbar$ expansion. In this sense, the time-dependent 
variational approach with the quasi-spin squeezed state 
gives the approximation including the higher order quantum fluctuations 
than $\hbar$ if any expansion is not applied.

\section{Summary}

In this paper, it has been shown that the exact ground state energy for the 
pairing model can be 
well recovered by using the time-dependent variational approach 
with the quasi-spin squeezed state. 
For this purpose, we treated the $su(2)$-algebraic model because 
its eigenvalue is known analytically. As a result, by taking into account 
of the dynamics in our quasi-spin squeezed state approach, the exact 
ground state energy can be reproduced up to the order of $1/\Omega$ 
under the small $|Y|^2$ expansion. 
Of course, the time evolution of a system governed by the pairing model 
Hamiltonian can be also investigated as is similar to that of the 
Lipkin model developed in Ref.\citen{TA04}. 
However, we do not repeat it because 
the result is almost same as that of the Lipkin model, which was reported in 
Ref.\citen{TA04}.

\section*{Acknowledgements} 
The authors would like to express their sincere thanks to Professor\break
M. Yamamura for valuable discussions. 
One of the authors (Y.T.) 
is partially supported by the Grants-in-Aid of the Scientific Research 
No.15740156 from the Ministry of Education, Culture, Sports, Science and 
Technology in Japan. 


\end{document}